\documentclass{article}

\usepackage[english]{babel}

\usepackage[letterpaper,top=2cm,bottom=2cm,left=3cm,right=3cm,marginparwidth=1.75cm]{geometry}

\usepackage{amsmath}
\usepackage{graphicx}
\usepackage[colorlinks=true, allcolors=blue]{hyperref}

\title{Extracting Spatial Interaction Patterns between Urban Road Networks and Mixed Functions}

\author{Huidan Xiao,Tao Yang}

\begin{document}
\maketitle

\begin{abstract}
In the field of urban planning, road network system planning is often the first step and the main purpose of urban planning is to create a spatial configuration of different functions such as residence, education, business, etc. Generally speaking, the more mixed the functions of an area has, the more possible its vitality may be. Therefore, in this article, we propose a new framework to study the specific spatial influence patterns of the overall structure and different sub-structures of road networks on the mixed functions. Taking road segment as the basic unit, we characterize mixed functions aggregation of road networks with the number of POIs categories within 100 meters around every road segment. At the same time, on the basis of centrality measurement in graph theory, we use 4 indexes to reflect the characteristics of the urban road network structure, including degree, closeness, betweenness, and length. We conduct our methods and experiments using the road networks and POI data within the 5th ring road of Beijing. The results demonstrate that urban road networks inherently influence the aggregation of urban mixed functions in various patterns and the patterns of road network sub-structures is also quite different. Our study shows that the higher the degree of the road network structure has, the more likely it will attract functions’ aggregation. It also reveals that diversified local degree will help gather urban functions. In addition to those, the analysis as well validates the importance of small-grids typology of urban road networks and the closeness to the center of cities.
\end{abstract}

\section{Introduction}

Urban mixed functions refer to the presence of two or more urban functions (e.g.,commercial, residential and business) in a certain space or time range in a city (ULI., 2003)\cite{1}, which is the driving force for the city to maintain vitality, diversity and charm. Jan Jacobs (1961)\cite{2} believes that city with diverse functions can help people to contact and communicate, increase the pleasant atmosphere and sense of security. A university town that mixes functions of education, research, business and residence is a typical example. In general, urban mixed functions is the result of the combined effects of factors such land use, spatial morphology, and functional layout. It is a spatial combination of different functions within a certain time range, and maintains complex relationships with each other.

With the development of new urbanism and smart growth, more and more studies on mixed use or functions have emerged (Grether and Mieszkowski, 1980\cite{3}; Geoghegan et al.,1997\cite{4}; Song and Knaap, 2004\cite{5}; Koster, 2012\cite{6}). One of the main common features of these research is that they most use the land use attribute of parcel or TAZ (Traffic Analysis Zone) to evaluate the degree to which how uses or functions are mixed. Whereas, because of their limitation in sampling number and accuracy in depicting the characteristics of mixed
functions (Yue et al., 2017\cite{7}), parcel-level or larger scale spatial land use data is more suitable for studying mixed functions at the regional level rather than within the city. However, the rapid advance in technologies such as location-based big data and ICT (Information and Communication Technology) make it possible for us to research mixed functions of cities in a fine-grained scale based on new data sources, especially POI data. For example, Zhou et al.(2016)\cite{8} studied the influence of urban mixed functions on energy consumption. Yue et al.(2017)\cite{7} compared the efficiency of different measurement in modelling relations between mixed functions and urban vibrancy. Feng et al. (2020)\cite{9} predicted housing prices by means of assessment of the accessibility of different functional facilities to the house. All of these literatures have shown that geographic big data can well describe the different characteristics of urban functions. So, in this study, we use the richness, which is the number of POI categories, as the measurement of mixed functions. Except for richness, there are also other indices for evaluating mixed functions such as entropy, density, and so on. Among all of them, entropy has been used most in various researches (Frank and Pivo, 1994\cite{10}). But it has been pointed out that entropy represents more about uncertainty rather than diversity (Jost 2006\cite{11}). Besides, Yue et al. (2017)\cite{7} has already found POI richness contributes more significantly to urban vibrancy than entropy. Furthermore, our focus in this research is to study the specific patterns of road network structure on mixed functions, so we take POI richness as a measure for representing urban mixed functions.

Meanwhile, as the basic skeleton that shapes urban spatial form and functional layout (Jacobs, 1961\cite{2}; Gehl, 2011\cite{12}), road network system planning is the foundation of urban planning. Studying how its feature influence mixed functions exactly will facilitate us to create a more logic configuration of urban functions, and plan a more livable and sustainable city. Hillier et al. (1993)\cite{13} demonstrated the influence of road network structure on pedestrian movement patterns combining cognitive psychology and graph theory. Afterwards, studies started to focus more on the attributes of urban road networks from the perspective of graph (Lu et al., 2018\cite{14}; Ahmadzai et al., 2019\cite{15}) and explore its relations with factors like accessibility (Batty, 2009\cite{16}), land use (Penn et al., 2004\cite{17}; Shen et al., 2013\cite{18}) and so on. Although the literature has shown insight for analyzing road networks, they rarely discuss the exact spatial interaction patterns between it and mixed functions within a city. As a result, the contribution of geographic big data such as POI has been greatly limited (Ratti, 2004\cite{19}; Geurs et al., 2012\cite{20}).The few current researches (Shen et al., 2016\cite{21}) only studied the correlation of these two and identified different urban function areas without tapping the compound pattern about how they interact with each other, which is just quite essential for urban planning. Since the planning of urban road networks tend to be the first step of urban planning in most cases and it also influences the planning of functional zones.

Therefore, in this article, we propose a new framework to extract exact interaction patterns of urban road networks and mixed functions spatially. We try to reveal the complex characteristics of urban road network structure and how it influences mixed functions in detail, comprehensively using graph theory and data mining algorithms. We first verify that the overall road network structure has an important influence on mixed functions, and then discuss the different patterns of sub-structures. Here, the overall structure refers to the network structure composed of all road segments in a city or the central area of a city. The sub-structure of refers to the network structure of different patterns formed by road segments,according to their different characteristics. All the sub-structures constitute the overall road network structure. There are three reasons why we divide the road networks into overall structure and sub-structures. First, patterns may differ in overall structure and sub-structures.Second, check if the features can well describe the characteristics of road networks. Third, in the field of urban planning, there are some kinds of road networks with special morphology such as radical and grids.The research is an exploratory and important step to study the compound interaction patterns of road network structure on mixed functions, enabling us creating a better spatial functional planning by optimizing the road network structure. The main contributions of this article can be summarized as the following points:

\begin{itemize}
\item We provide a new perspective and effective framework for quantifying how the urban road networks influence urban mixed functions.
\item We successfully divided the road network into an overall structure and 4 sub- structures based on our proposed indices for evaluating road networks.

\item We find the impact patterns of the overall structure and sub-structures on mixed functions in multiple dimensions in detail.
\end{itemize}

\section{Problem Formulation}

\subsection{Mixed Functions}

Definition 1:(Richness) Richness refers to the number of types of POIs within buffered area of each road segment and is represented as R=S, where S represents the number of types of POIs. The greater the value of richness is, the richer the functions’ types are around a road segment.
\subsection{Urban Road Network Structure Measurement}
We mainly draw on the centrality measurement in graph theory and take every road 85 segment as a node in a graph. The exact definitions and formulas are as follows:

Definition 2:(Degree) Degree refers to the number of nodes directly connected to node i in the connected network and is represented as $C_{i}^{c}$=$K$,where $C_{i}^{c}$ resents the 
degree of 88 node i and K is the number of nodes direct connected to node i. We use this indicator as follows to represent the local connectivity of each road segment of the urban road networks

Definition 3:(Closeness) Closeness refers to the reciprocal of the sum of the distances 91 between node i and all the other nodes in the network, reflecting the proximity of a node to 92 other nodes. The higher the closeness, the shorter the path from this node to all other nodes 93 and more likely to be the geometric center of this road networks. Hence, we use this indicator 94 here to represent the proximity of node i to the geometric center of the urban road networks, 95 as it is calculated as shown in Eq. 1:

\begin{equation} \label{}
C_{i}^{c}=\sum_{j=1, i \neq j}^{N} d_{i j},(1)\;.
\end{equation}
where $C_{i}^{c}$ represents the closeness of node i and N represents the total number of nodes in the networks. $d_{i j}$ is the  distance of the shortest path between node i and node j.

Definition 4:(Betweenness) Betweenness refers to the number of shortest paths of any two nodes in the network through a node. The greater the betweenness of a node, the more the shortest paths through this node and it is calculated as shown in Eq. 2:
\begin{equation} \label{}
C_{i}^{b}=\sum_{j=1, k=, 1 j \neq k \neq i}^{N} \frac{n_{j k}(i)}{n_{j k}},(2)\;.
\end{equation}
where $C_{i}^{b}$ represents the betweenness of node i and N represents the total number of nodes in the networks.$n_{j k}$
represents the number of shortest paths from node j to node k,
and respectively $n_{j k}(i)$ represents the number of shortest paths which pass-through node i.

Definition 5:(Length) Length refers to the physical length of every road segment in the urban road networks, and is denoted as $C_{i}^{l}=Dist(i)$ where $C_{i}^{l}$ represents the length of every road segment.

\section{Methodology}
This research uses road segment as the basic study unit, aiming to find the specific spatial interaction patterns between urban mixed functions and urban road networks structure, and the flowchart of our proposed method is shown in Fig. 1.We first perform a preliminary data cleaning on the original OSM road network data, and convert it into segmented road map. Then, we transform Gaode POIs into a high-dimensional vector and obtain the final characteristic vector of compound functions of every road segment. Next, we construct the characteristic vector of structural characteristics of each road segment mainly based on graph's centrality measures. Finally, we explore spatial interaction patterns of urban road networks and mixed functions using ensemble methods, clustering and statistical analysis.

\subsection{Characterize vectors of Mixed Functions Aggregation}
In this study, we use the richness of POIs to reflect mixed functions. In order to integrate richness with road networks based on road segment, we take the following steps:
\begin{enumerate}
\item Select 47 types of POIs related to urban functions based on expert knowledge for quantitative statistics.
\item Buffer the road segment at an appropriate radius and then calculate the richness of each road segment (Fig. 2(c)).
\item Characterize richness vectors of each road segment.
\end{enumerate}

\begin{figure}
    \centering
    \includegraphics[width=0.95\textwidth]{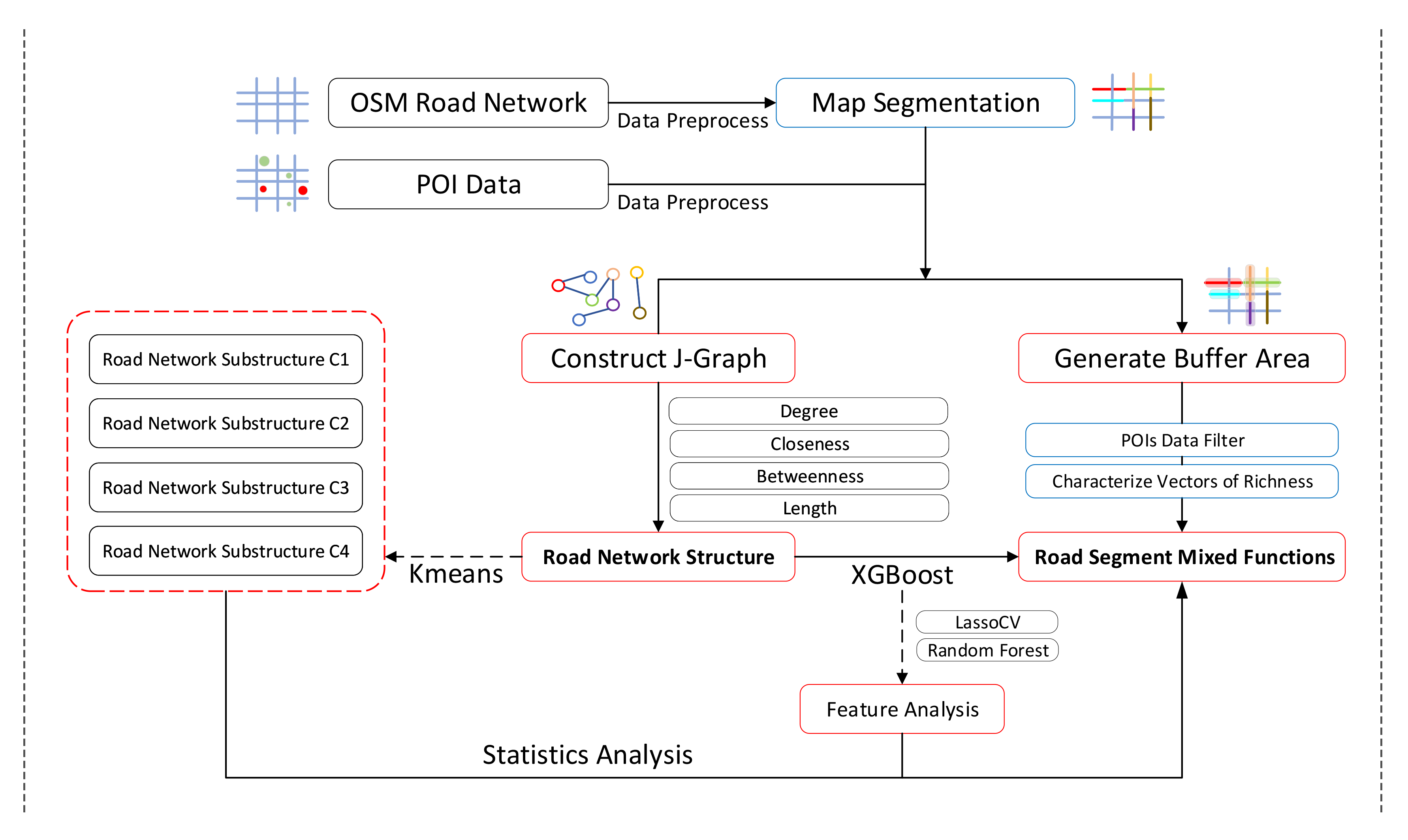}
    \caption{Flowchart}
    \label{fig:galaxy}
\end{figure}

\subsection{Characterize Road Network Structure}
The urban road network structure expressed as a graph is G = (V, E), where V represents the set of nodes and E represents the set of edges. We first process OSM Data to obtain the segmented road map, and then extract the roads of one to three levels as the raw data for creating J-graph as shown in Fig. 2(b). Next, we use every road segment as vertex and links between road segments as the edges to build the needed graph structure (Fig. 2(d)). The process to transfer urban road networks to segmented graph is shown in Fig. 2.

\begin{figure}[hp]%调节图片位置，h：浮动；t：顶部；b:底部；p：当前位置
    \centering
    \includegraphics[width=0.95\textwidth]{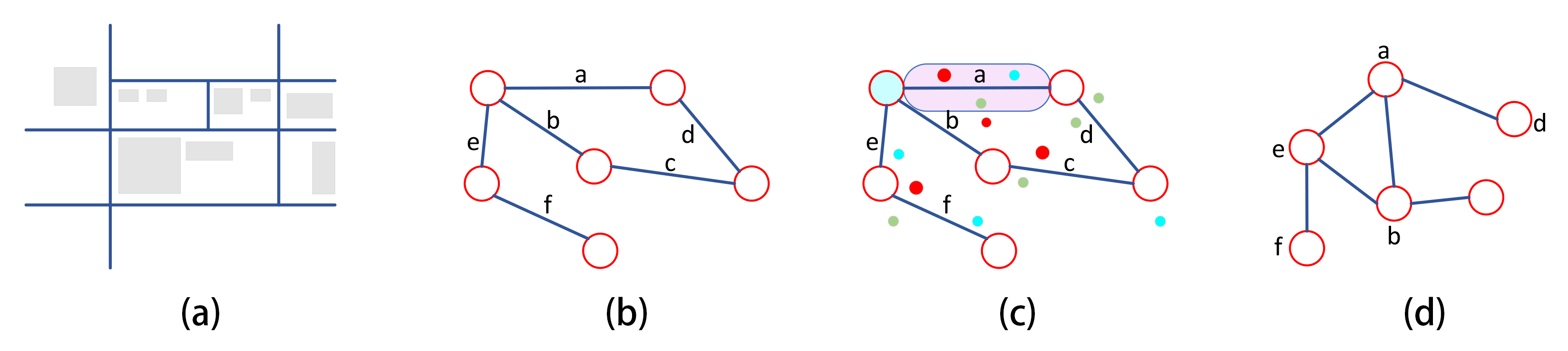}
    \caption{Illustration of constructing J-graph and vectors of richness}
    \label{fig:galaxy}
\end{figure}

After the J-graph is constructed, we calculate degree, betweenness, and closeness of each road segment according to the centrality measures in graph theory. Here we use central 
measures because they are widely used in physical research world (Ahmadzai et al.,2018\cite{15}; Liu et al.,2019\cite{22}; Wang et al., 2020\cite{23}; Lu et al., 2018\cite{14}). Wang, et al (2020)\cite{23} use road networks to study ride-sharing accessibility. Yang et al (2019)\cite{24} analyze dockless sharing patterns based on graph.Besides, they also play an important role in the research field of space syntax, not only focusing on road networks form itself but also try to correlate it with other socioeconomic problems (Penn et al.,2004\cite{17}; Karimi, 2012\cite{25}; Hillier, 2007\cite{26}). In addition, we also calculate the length of each road segment as a measure of the structure of urban roads. Because the more compact the road networks are, the more likely the short streets are to achieve diversity generally.

\subsection{Identification of Spatial Interaction Patterns}
Because of its advantage in non-linear data fitting (Fan et al.,2018; Hadri et al.,2018\cite{27};Reynaldo et al.,2019\cite{28}), we choose xgboost to fit the relationship and distinguish different feature's importance. After constructing the characteristic vectors of urban road networks and function composite of each road segment, we then use xgboost to extract the relationship between these two factors. xgboost is an ensemble algorithm which builds and combines multiple weak learners to complete a difficult learning task. The core idea is that the basic model of each training is fitted to the residuals of the last trained basic model to continuously reduce it in the principle of addition strategy. And the basic model is generally CART classification or regression tree. The t-th prediction result of xgboost is generally expressed as: 
\begin{equation} \label{}
\hat{y}_{i}=\sum_{k=1}^{t} f_{k}\left(x_{i}\right)=\hat{y}_{i}^{(t-1)}+f_{t}\left(x_{i}\right),(3)\;.
\end{equation}
where $k$ represents the total number of trees and $k$ is the kth tree. $x_{i}$ presents the input variable. To achieve a bias-variance tradeoff while optimizing model performance and computational speed, the objective function of xgboost are usually added with regularization terms and the objective function at step $t$ is as follows:

\begin{equation} \label{}
obj^{(t)}=\sum_{i=1}^{n} l\left(y_{i}, \hat{y}_{i}^{(t-1)}+f_{t}\left(x_{i}\right)\right)+\Omega\left(f_{t}\right),(4)\;.
\end{equation}
where the first term is the loss function and the second term is the regularizer, which is
defined as:
\begin{equation} \label{}
\Omega(f)=\gamma T+\frac{1}{2} \lambda\|w\|^{2},(5)\;.
\end{equation}
where $\gamma$ is the complexity cost to future partition the node, T is the number of leaves, $\lambda$ is the hyperparameter, and $\frac{1}{2} \lambda\|w\|^{2}$ is is the scores of leaves to measure the structure of the tree.
Usually, for the purpose of fast calculation, we can approximate it by means of second order of Taylor expansion as shown in Eq. 6:
\begin{equation} \label{}
obj^{(t)} \cong \sum_{i=1}^{n}\left[l\left(y_{i}, \hat{y}_{i}^{(t-1)}\right)+g_{i} f_{t}\left(x_{\mathrm{i}}\right)+\frac{1}{2} h_{i} f_{t}^{2}\left(x_{i}\right)\right]+\Omega\left(f_{t}\right),(6)\;.
\end{equation}

where $g_{i}=\partial \widehat{y}^{(t-1)} l\left(y_{i}, \widehat{y}^{(t-1)}\right)$ and $h_{i}=\partial^{2} \hat{y}^{(t-1)} l\left(y_{i}, \hat{y}(t-1)\right)$ are the first and second order radiant of the loss function. We keep the whole training process repeatedly until the stopping criterion are met, and more detailed explanation about xgboost can be found in Chen and Guestrin (2016)\cite{29}. Then, we evaluate the importance of different features to mixed functions based on our model.

Next, K-means is adopted to classify road networks for the purpose of further analyzing the effectiveness of features to mixed functions in different road network sub-structures.
Because of its advantage in fast calculation and good performance, K-means has been widely used in different researches (Luo et al.,2011\cite{30}; Hu et al.2008\cite{31}). We cluster the road networks using the four features as vector dimensions, and our loss function to be optimized is as shown in Eq. 7:

\begin{equation} \label{}
E=\sum_{i=1}^{k} \sum_{x \in c_{i}}\left\|x-u_{i}\right\|^{2},(7)\;.
\end{equation}

where $u_{i}=\frac{1}{\left|C_{i}\right|} \sum_{x \in C_{i}} x$ is the mean vector of cluster $C_{i}$. As an iterative optimized clustering algorithm, the main steps of K-means is outlined as following:

\begin{figure}[ht]%调节图片位置，h：浮动；t：顶部；b:底部；p：当前位置
    \centering
    \includegraphics[width=0.95\textwidth]{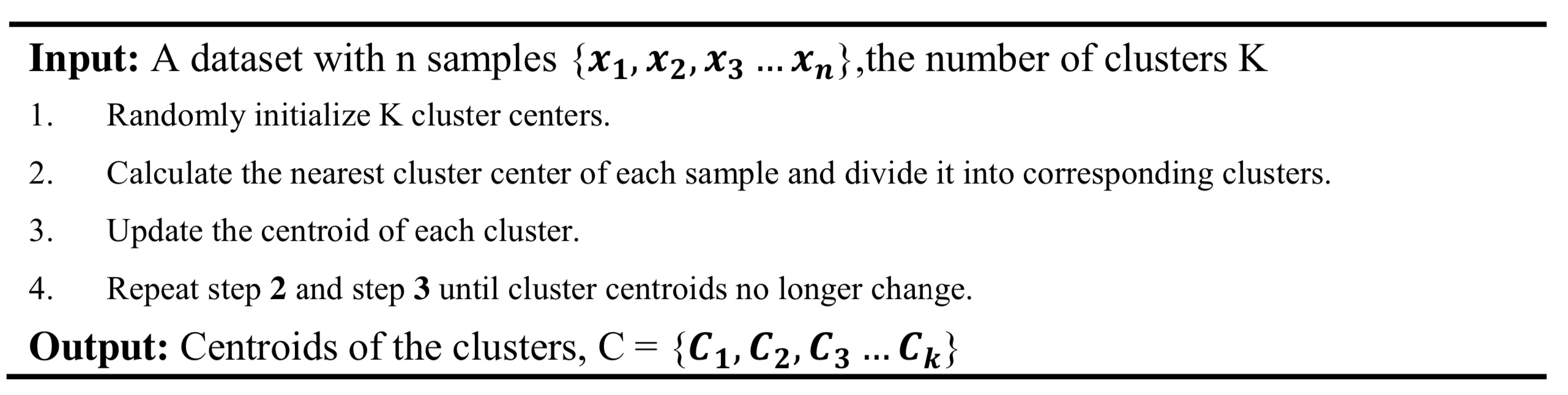}
    \caption{k-means algorithm}
    \label{fig:galaxy}
\end{figure}

An obvious shortcoming of K-means is the uncertainty of the clustering results, so we apply the elbow method to select the best number of clusters. Then, we combine our analysis with xgboost and clustering results to dig out more of the latent factors of the spatial interaction patterns of urban road networks and urban mixed functions. In the next section, we will talk about our analysis in detail.

\section{Experiments and Results}
\subsection{Study area and Data}
\textbf{(1)OSM Road Network Data}
We study the area within 5th ring road of Beijing, which is a major urban built-up area in Beijing with an area of about 670 square kilometers. We first download the road networks of Beijing from OSM open data source of 2019 and then process it to the final road data as shown in Fig. 4(a) by cleaning and filtering it manually.

\begin{figure}[ht]%调节图片位置，h：浮动；t：顶部；b:底部；p：当前位置
    \centering
    \includegraphics[width=0.95\textwidth]{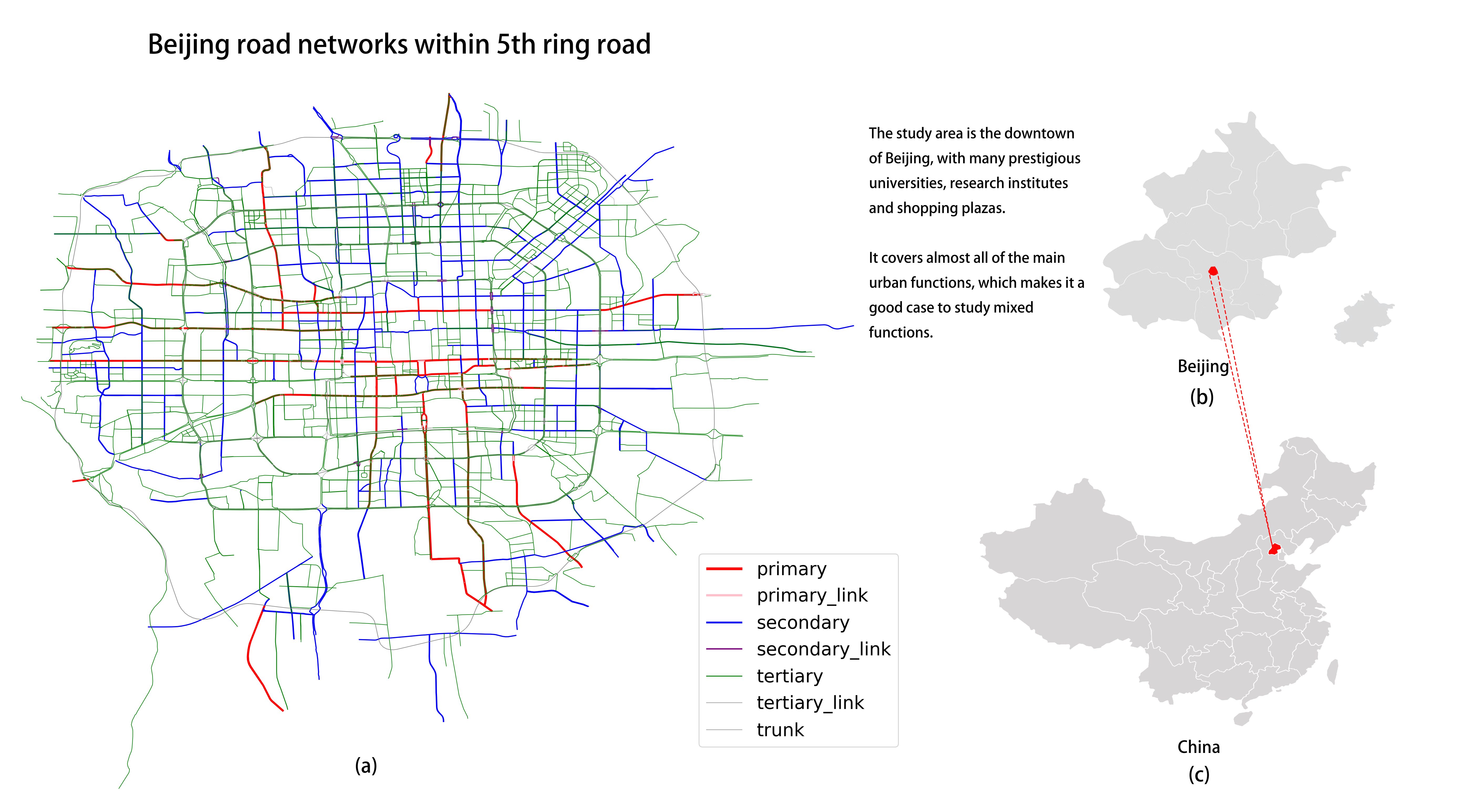}
    \caption{(a) Road network within 5th ring road of Beijing from OSM; (b)Beijing; (c)China.}
    \label{fig:galaxy}
\end{figure}

\begin{figure}[ht]%调节图片位置，h：浮动；t：顶部；b:底部；p：当前位置
    \centering
    \includegraphics[width=0.95\textwidth]{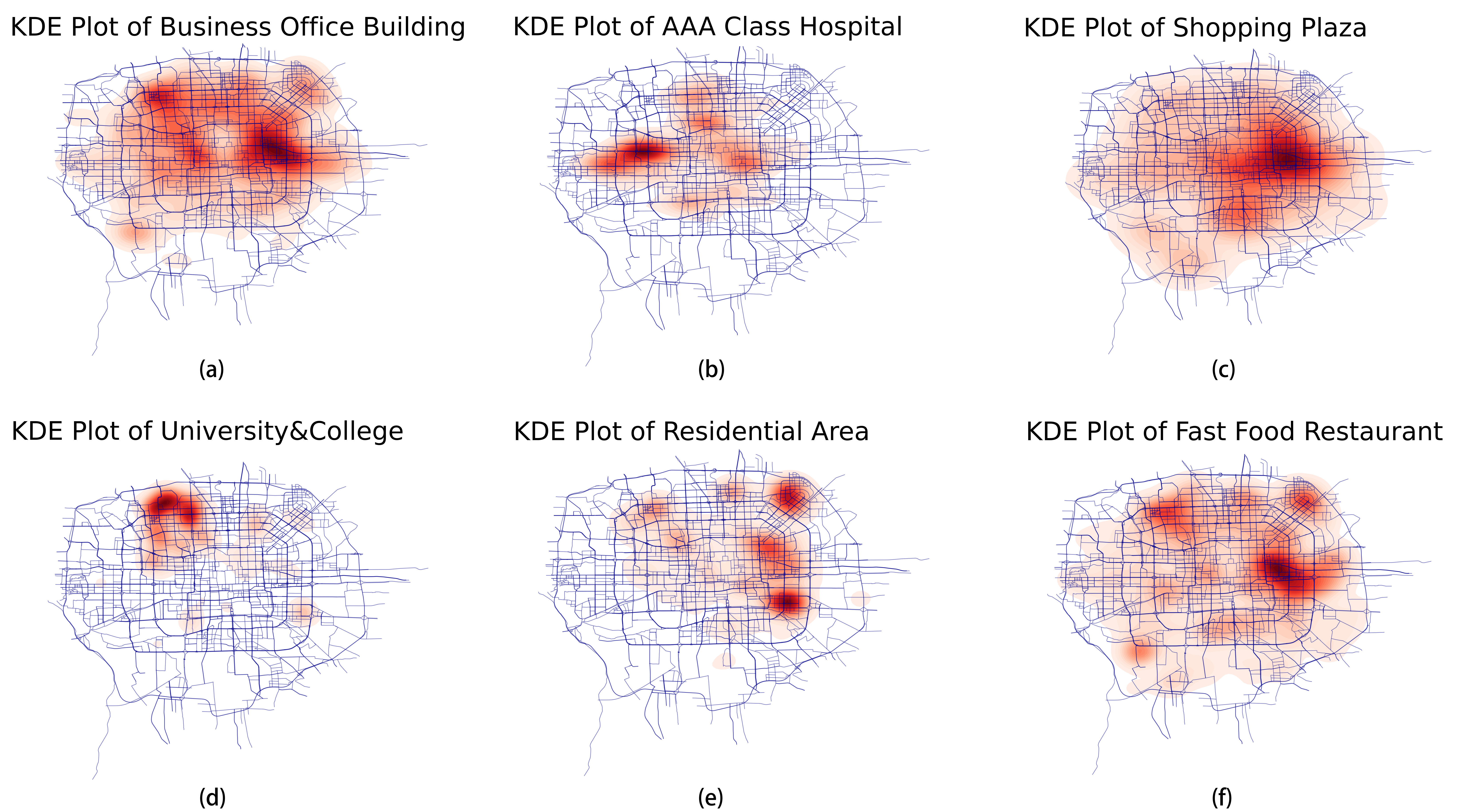}
    \caption{(a) Fig. 5. Intensity of Functions (a)Business Office Building; (b)AAA Class Hospital; (c)Shopping Plaza; (d)University \& College; (e)Residential Area; (f)Fast Food Restaurant..}
    \label{fig:galaxy}
\end{figure}

\textbf{(2)POIs Data}

POIs data comes from Gaode Map, a location-based service company similar to Google Maps. The Beijing POIs dataset in 2018 we access has 23 first-class types, 240 second-class types, 3125 third-class types in total. Because of the fineness of the functional description of the third-class of POIs data and the big quantity of its number, we finally select 47 third-class types of POIs to evaluate functions’ richness in our experiment, and their statistics are shown in Table 1 and Table 2:

\begin{table}[p]
\centering
\includegraphics[width=0.95\textwidth]{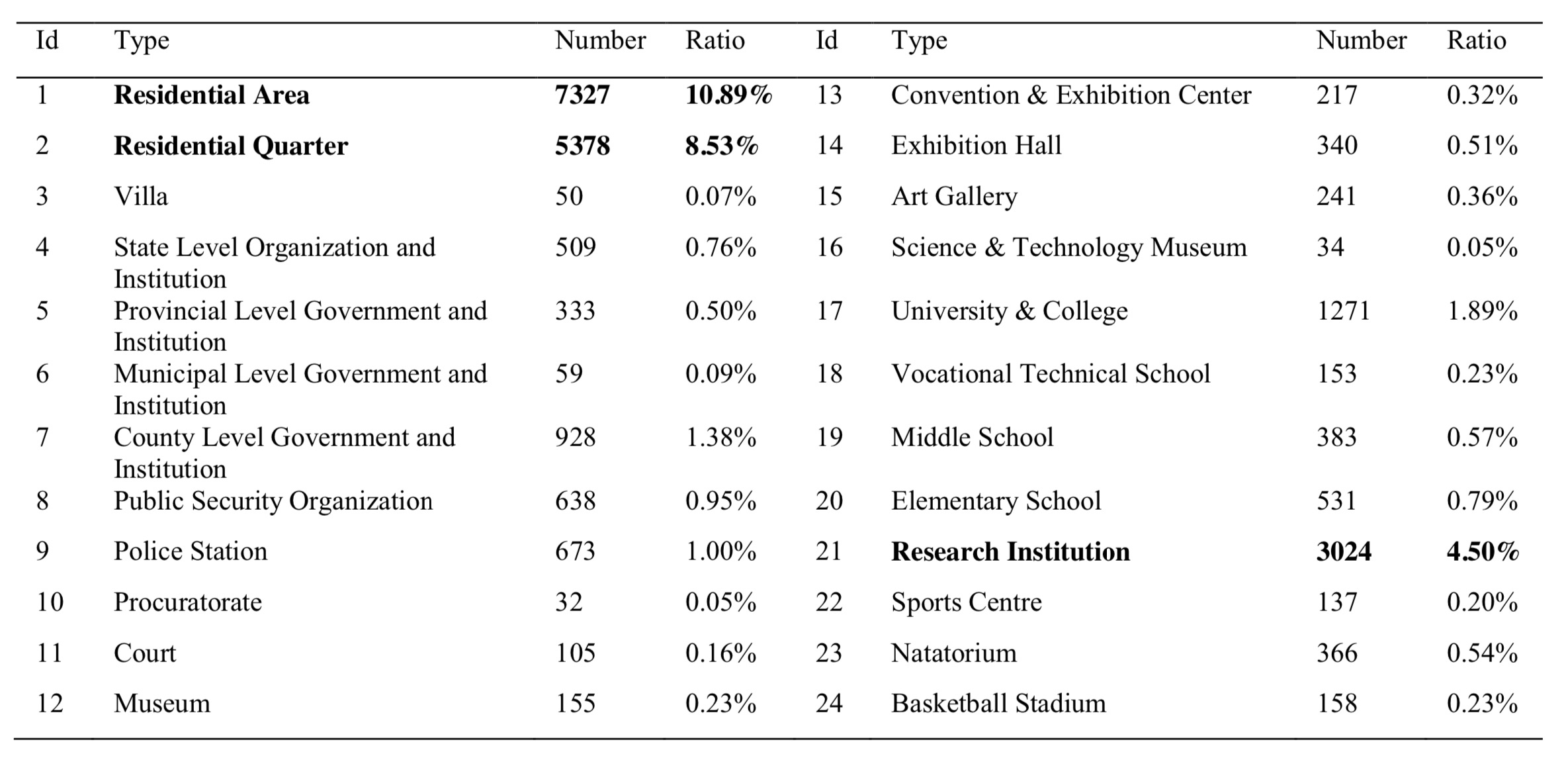}
\caption{\label{tab:widgets}Statistics of POI data within 5th ring road of Beijing-Part1.}
\end{table}

\begin{table}[p]
\centering
\includegraphics[width=0.95\textwidth]{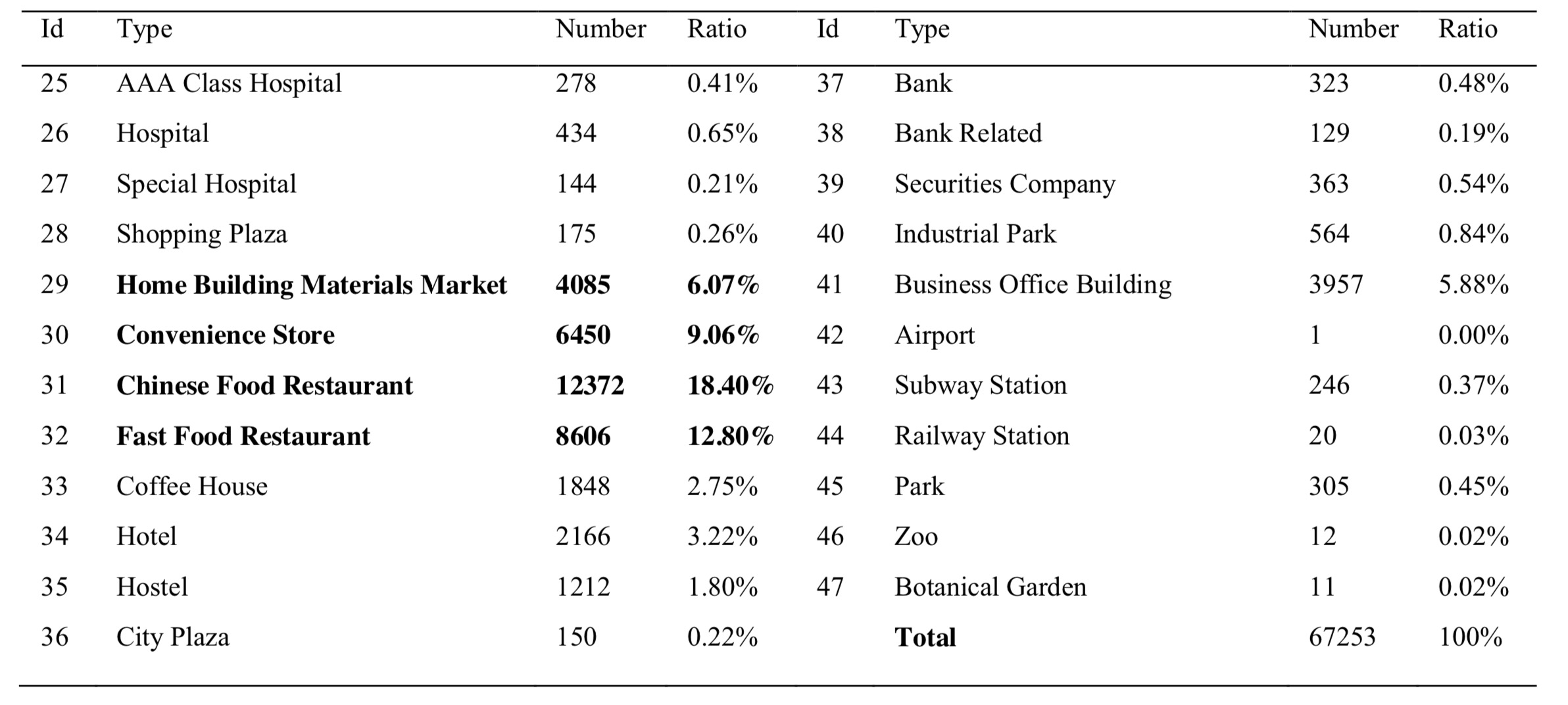}
\caption{\label{tab:widgets}Statistics of POI data within 5th ring road of Beijing-Part2.}
\end{table}

From Table 1 and Table 2, we can see that the types of POIs which accounts for more than 10 within the 5th ring area are Chinese Food restaurants, Fast Food Restaurant, and Residential Area. The types that account for between 5 and 10 are Convenience Store, Residential Quarter, Home Building Materials Market and Business Office Building. Among them, Fast Food Restaurants and Convenience Store belong to small-scale living service facilities, which mainly depend on the existence of residential areas, and Business Office Buildings are generally the main elements of the commercial area, in which there are also a large number of fast food restaurants and Chinese restaurants around. In addition, the number of POIs of research institutions also ranks eighth in the total number of 47 types of POI types within the 5th ring road, accounting for 4.5. Therefore, from the statistics of POIs, we can see the functions within Beijing's 5th ring road are mainly residential,commercial and scientific research.

Fig. 5 shows us the examples of function density estimation of a single POI type spatially, for example, Fig. 5(c) shows that large shopping malls are mainly distributed in the east, while Fig. 5(d) shows that universities and colleges are mainly centered in the northwest corner of the 5th ring area. It is apparent that different functions are often aggregated in different places of the city in some kind of pattern. So, we will further discuss the patterns in detail in next section.

\subsection{Results and Analysis}
\subsubsection{Validation of the effectiveness of XGBoost}
In order to better explain the influence patterns of the overall road network structure on functions’ aggregation and difference of features, we used lasso regression and random forest as baseline methods. Table 3 and Table 4 show the corresponding R-squared, RMSE (Root Mean Square Error) and MAE (Mean Absolute Error) of different models on train data and test data respectively. R- squared is a statistical measure for representing the proportion of the variance of a dependent variable which are explained by independent variables. The bigger the R-squared is, the better the model performs. RMSE is the standard deviation of the residual for measuring how the predicted values are concentrated around the true values. MAE is the mean error between predicted values and true values. The smaller these 2 indices are, the better the model performs.

From Table 3 and Table 4, xgboost performs best on the training dataset, and random forest has best performance on the test dataset when the parameter settings are close. As we can see, xgboost has the greatest R-squared on training dataset, reaching around 0.7 while lasso regression is only around 0.18. However, in case of the overfitting problem, so we also check the performance on test dataset. It turns out that random forest regression now performs best. But we can see there is only a little difference of R-squared about 0.06 between xgboost and random forest. Obviously, xgboost plays best in characterizing relationship between road network structure’s feature and mixed functions’ aggregation.

Fig. 6 shows the ranking of the importance of road network features under different evaluation criteria. From the perspective of feature weight, the most important feature is betweenness, that is to say feature of betweenness has the biggest number of splitting nodes in all decision trees. From the perspective of coverage, the most important feature is degree, which means that degree covers most samples when splitting nodes in all decision trees. From the perspective of coverage, feature of length has the most gain among features in all decision trees. Clearly, features have different importance when used with different methods, indicating the complexity of influence of a single road network structure’s feature on functions’ aggregation.

Fig. 8 shows the spatial distribution of different feature of the road network structure. From the Fig. 8, we can also see that all the features have a certain kind of consistency with richness (Fig. 7) spatially. So, in the next section, we will further explore the specific mode of each feature about how they affect richness exactly.

\begin{table}[ht]
\centering
\includegraphics[width=0.95\textwidth]{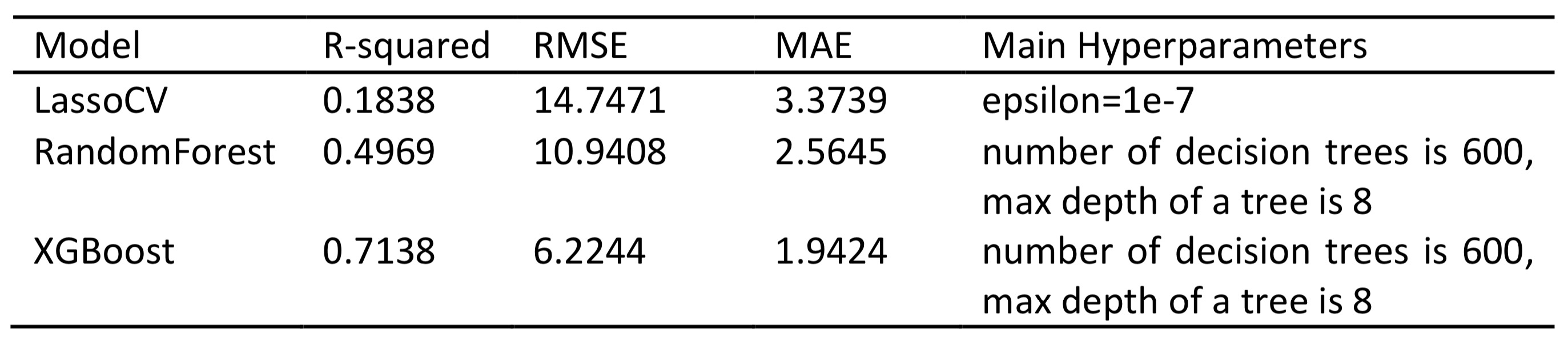}
\caption{\label{tab:widgets}Xgboost and baseline methods’ performance on train data}
\end{table}

\begin{table}[ht]
\centering
\includegraphics[width=0.95\textwidth]{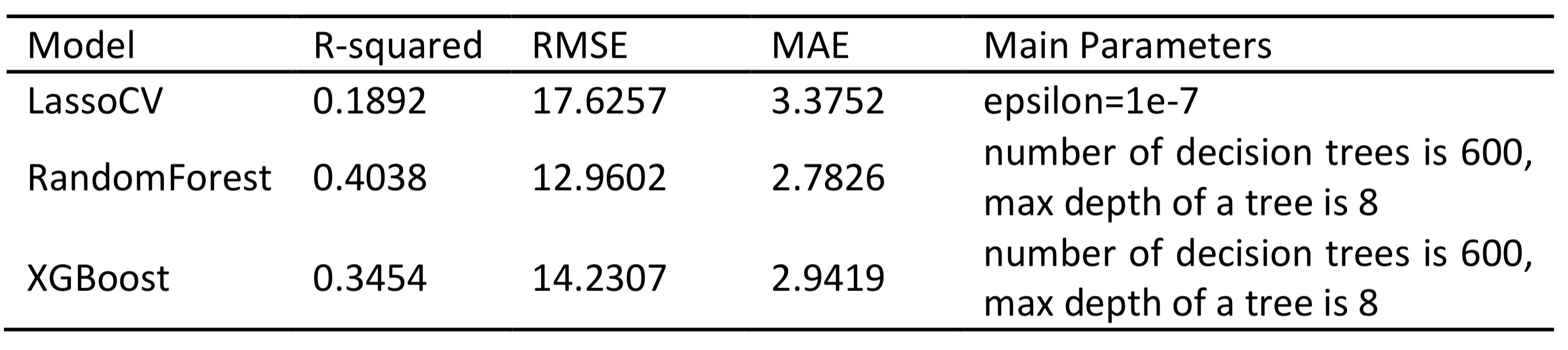}
\caption{\label{tab:widgets}Xgboost and baseline methods’ performance on test data}
\end{table}

\begin{figure}[p]%调节图片位置，h：浮动；t：顶部；b:底部；p：当前位置
    \centering
    \includegraphics[width=0.95\textwidth]{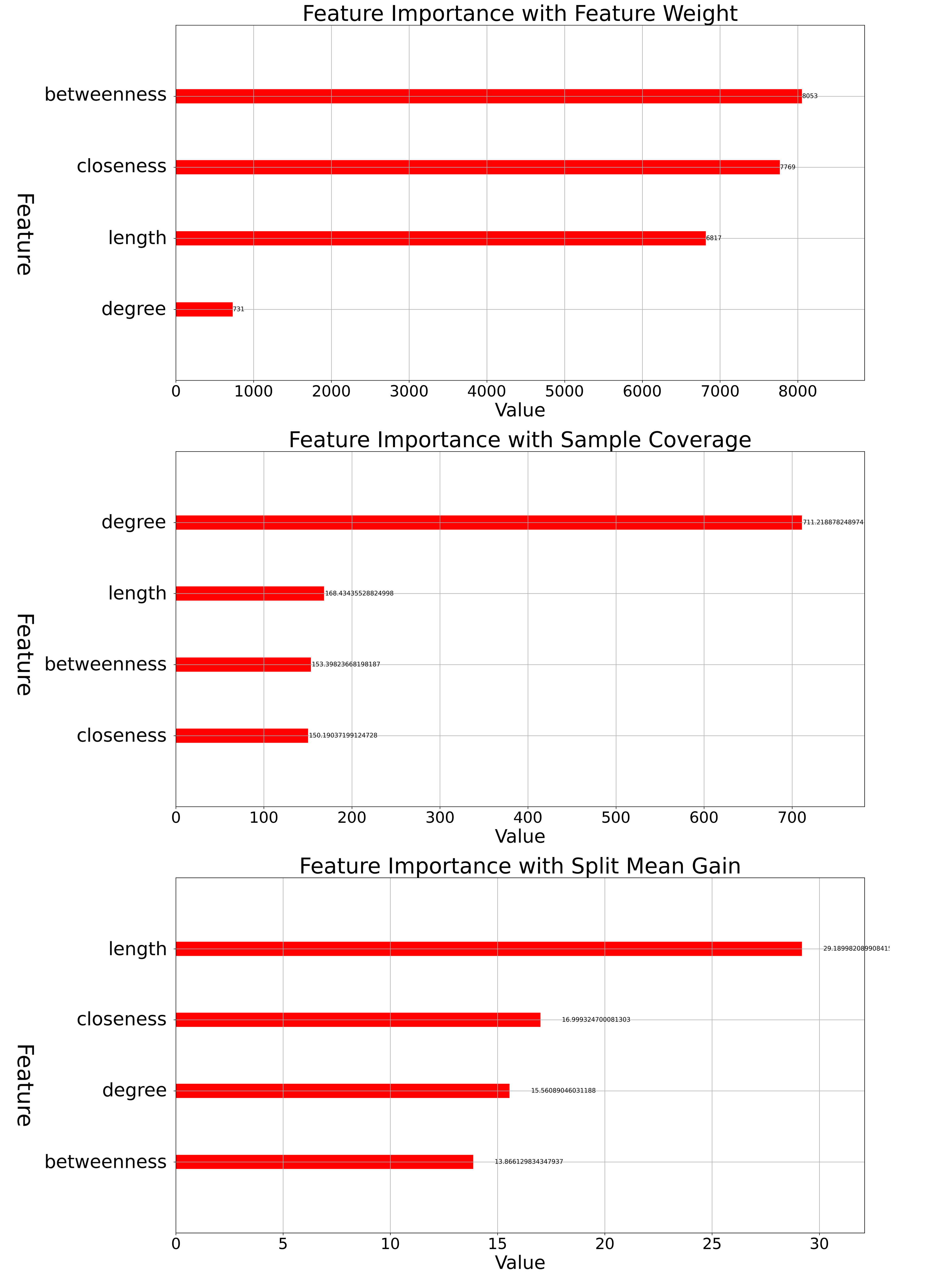}
    \caption{Feature importance with 3 methods of xgboost}
    \label{fig:galaxy}
\end{figure}

\begin{figure}[p]%调节图片位置，h：浮动；t：顶部；b:底部；p：当前位置
    \centering
    \includegraphics[width=0.95\textwidth]{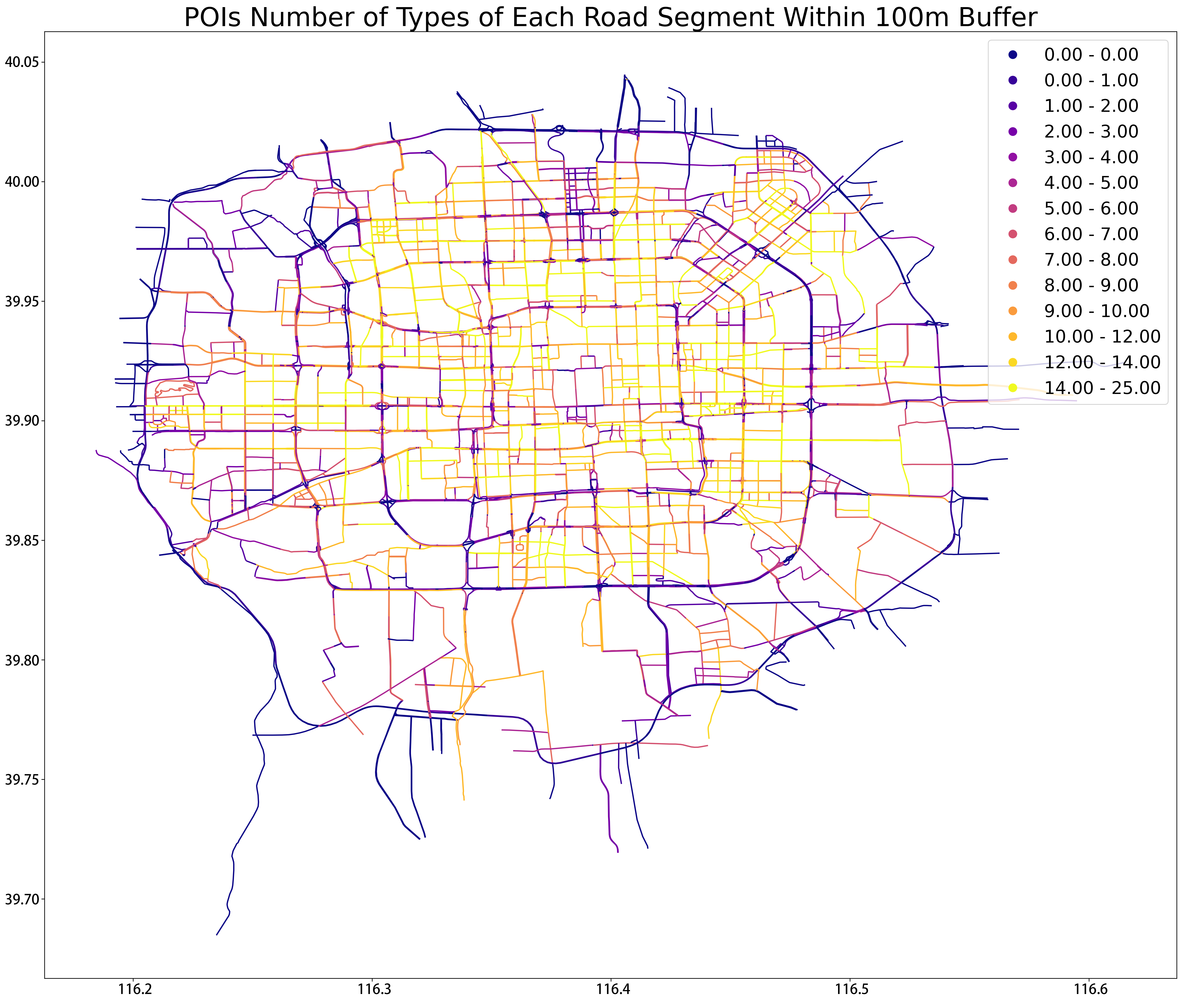}
    \caption{Spatial distribution of richness of road networks}
    \label{fig:galaxy}
\end{figure}

\begin{figure}[p]%调节图片位置，h：浮动；t：顶部；b:底部；p：当前位置
    \centering
    \includegraphics[width=0.95\textwidth]{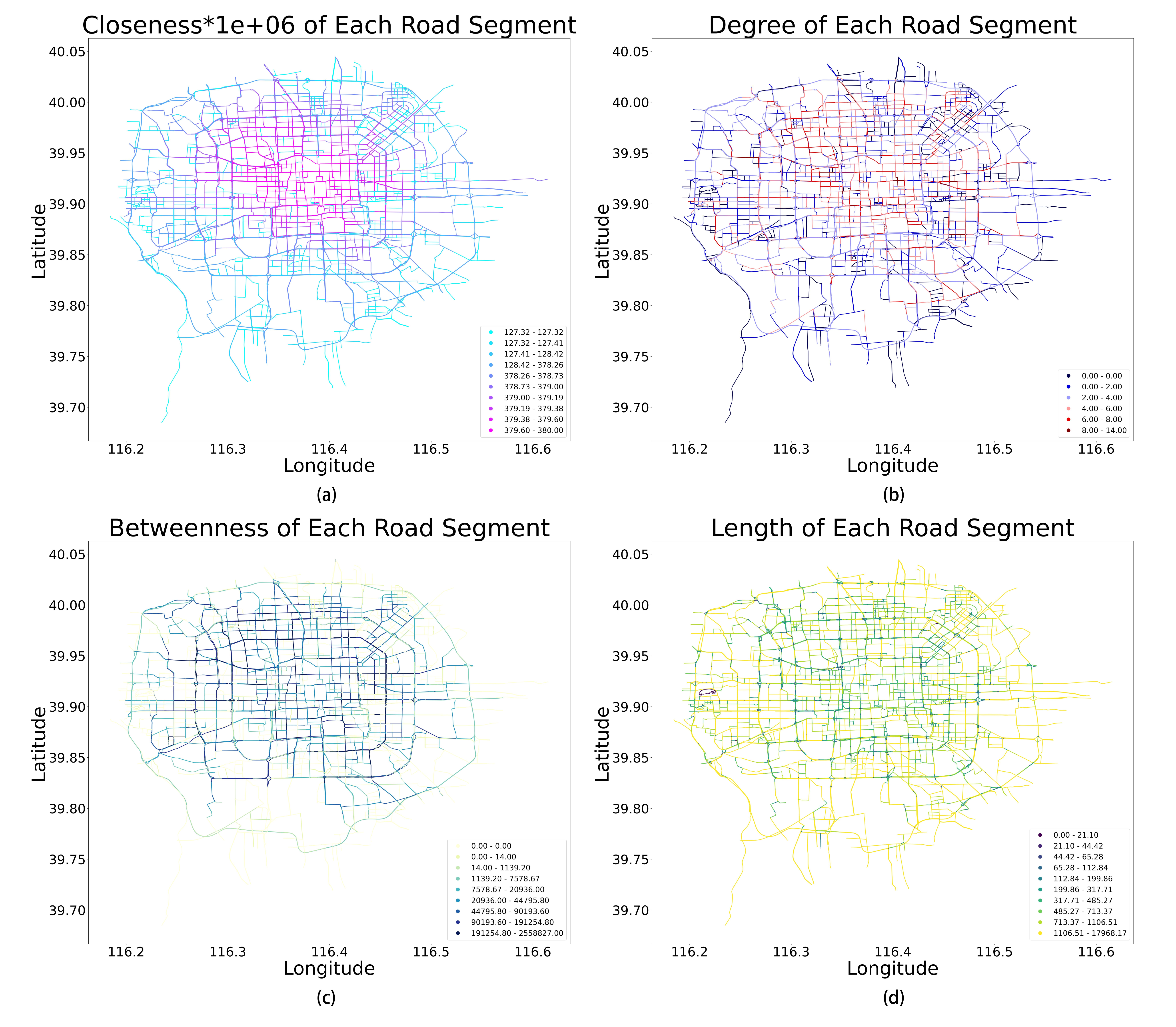}
    \caption{Spatial illustration of features of experimental road networks}
    \label{fig:galaxy}
\end{figure}

\subsubsection{Relations between four features of the overall road network structure and mixed functions}

After verifying the significant importance of the road network structure, with the aim of verifying how each feature influence mixed functions, we now further explore the relationship between richness and degree, betweenness, closeness and length respectively by drawing their confidence interval plot as shown in Fig. 9. Taking Fig. 9(b) as an example, when we draw the relationship between betweenness and richness, we first sort all the road segments according to the value of betweenness, and then divide the data into 10 groups in turn, take the average betweennes of these 10 data points as the independent variable and the average value of the corresponding richness as the dependent variable. Then, the blue line represents the change relationship between the dependent variable and independent variable. The gray area is the confidence interval of one standard deviation from the dependent variable's mean.Similarly, Fig. 9(c) and Fig. 9(d) are drawn in the same way as Fig. 9(b).Given that degree itself is discrete while other features are continuous, so when we draw the relationship between degree and richness, we just take every degree value as dependent variable and the average value of the corresponding richness as the independent variable, then we draw it similarly like the other 3 features.

First of all, Fig. 9 first shows that features which have a clear visible pattern are degree and closeness. There is a clear trend that the richness of each road segment increases as the degree increases from 4 to 6. Specifically, we can see that the richness increases rapidly as degree increases from 4 to 6 and keep a relatively stable state in a range of 6 to 10 of degree. Degree here we refer to local connectivity and it actually represents the number of other road segments directly connected to each road segment. In the point of view of topology, when degree is greater than or equal to 6, the morphology of the road networks will be very close to grids as shown in Fig.10,which explicitly depicts the evolution process of changes of different urban road networks’ morphology caused by degree. When the degree is greater than 10, in terms of the morphology of the road network, the road segments are either to be around the intersection of complex traffic nodes, which might be mixed with other functional areas like shopping plaza thus attract other functions’ aggregation and have a large value of richness. In general, the richness will increase with the increase of the degree and the grids will promote mixed functions great.

\begin{figure}[ht]%调节图片位置，h：浮动；t：顶部；b:底部；p：当前位置
    \centering
    \includegraphics[width=0.95\textwidth]{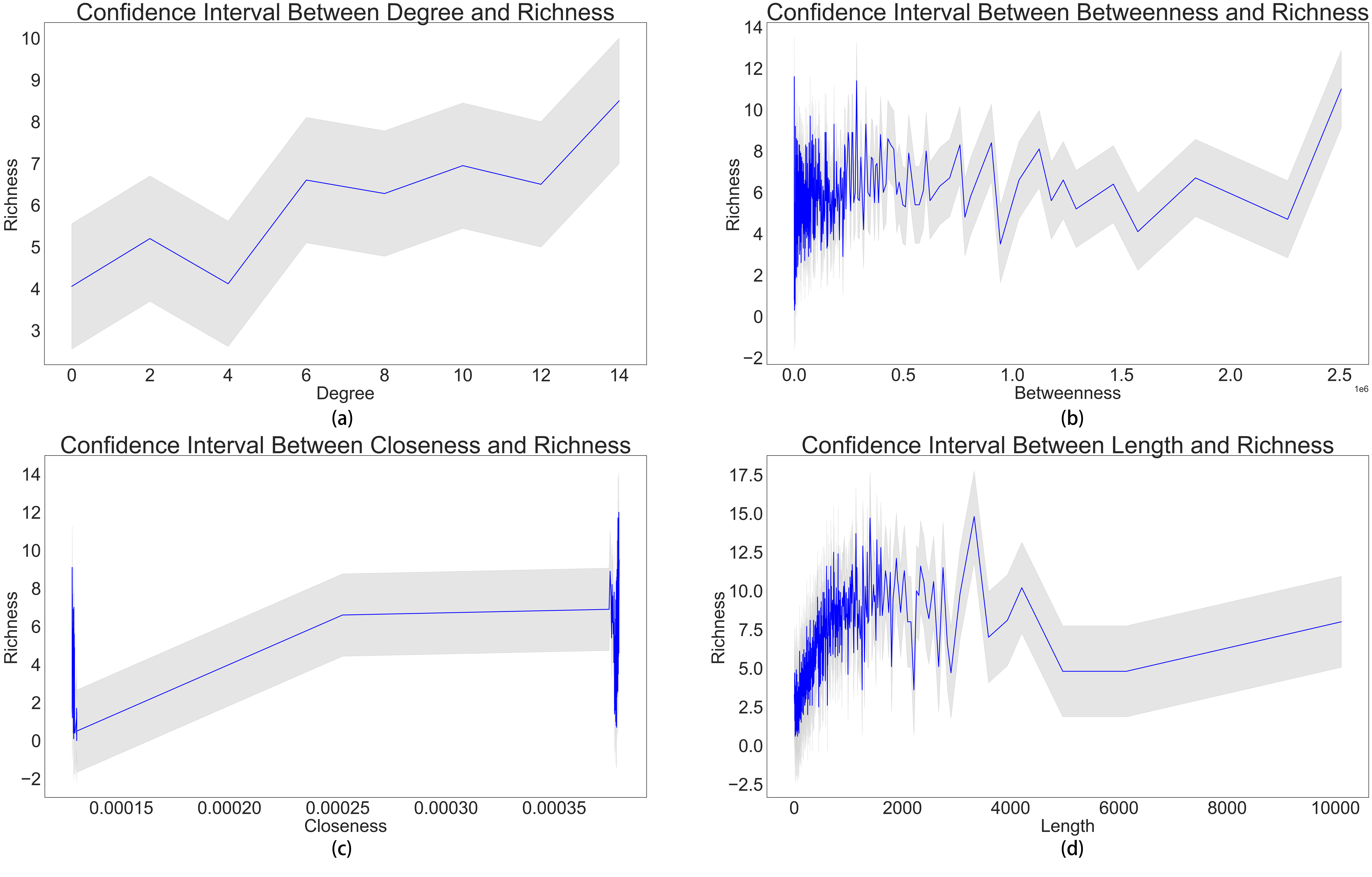}
    \caption{Confidence Interval between richness of 4 features of 4 road segments}
    \label{fig:galaxy}
\end{figure}

\begin{figure}[ht]%调节图片位置，h：浮动；t：顶部；b:底部；p：当前位置
    \centering
    \includegraphics[width=0.95\textwidth]{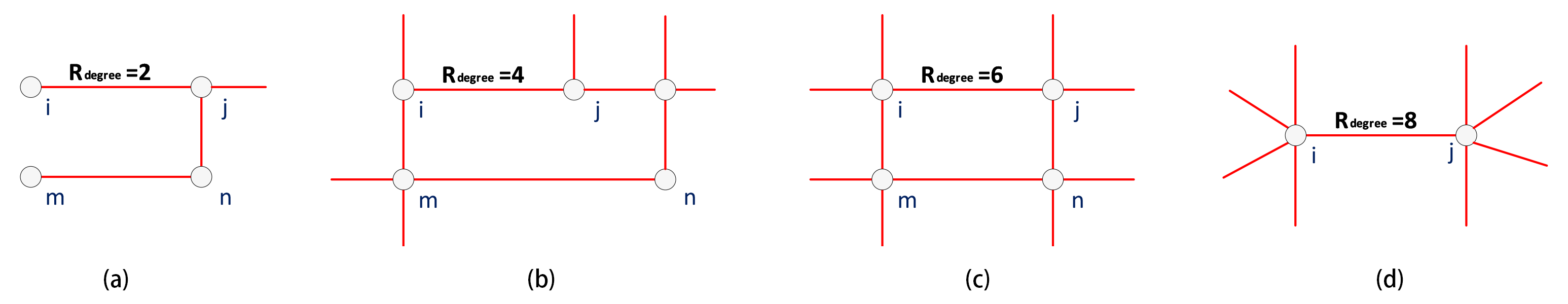}
    \caption{Road Network morphology of different values of degree}
    \label{fig:galaxy}
\end{figure}

Secondly, for closeness, Fig. 9(b) shows the obvious distribution around the minimum and maximum near the two ends of closeness. Given that closeness represents the reciprocal of the sum of the distances of the shortest paths of each road segment from all other road segments. So, the greater the closeness is, the closer the road segment is to the center of the road networks. Therefore, we first divide the roads into internal roads and external roads, then compare the richness of quantiles of these two kinds of road segments in the whole road network structure as shown in Fig. 11.
\begin{figure}[ht]%调节图片位置，h：浮动；t：顶部；b:底部；p：当前位置
    \centering
    \includegraphics[width=0.95\textwidth]{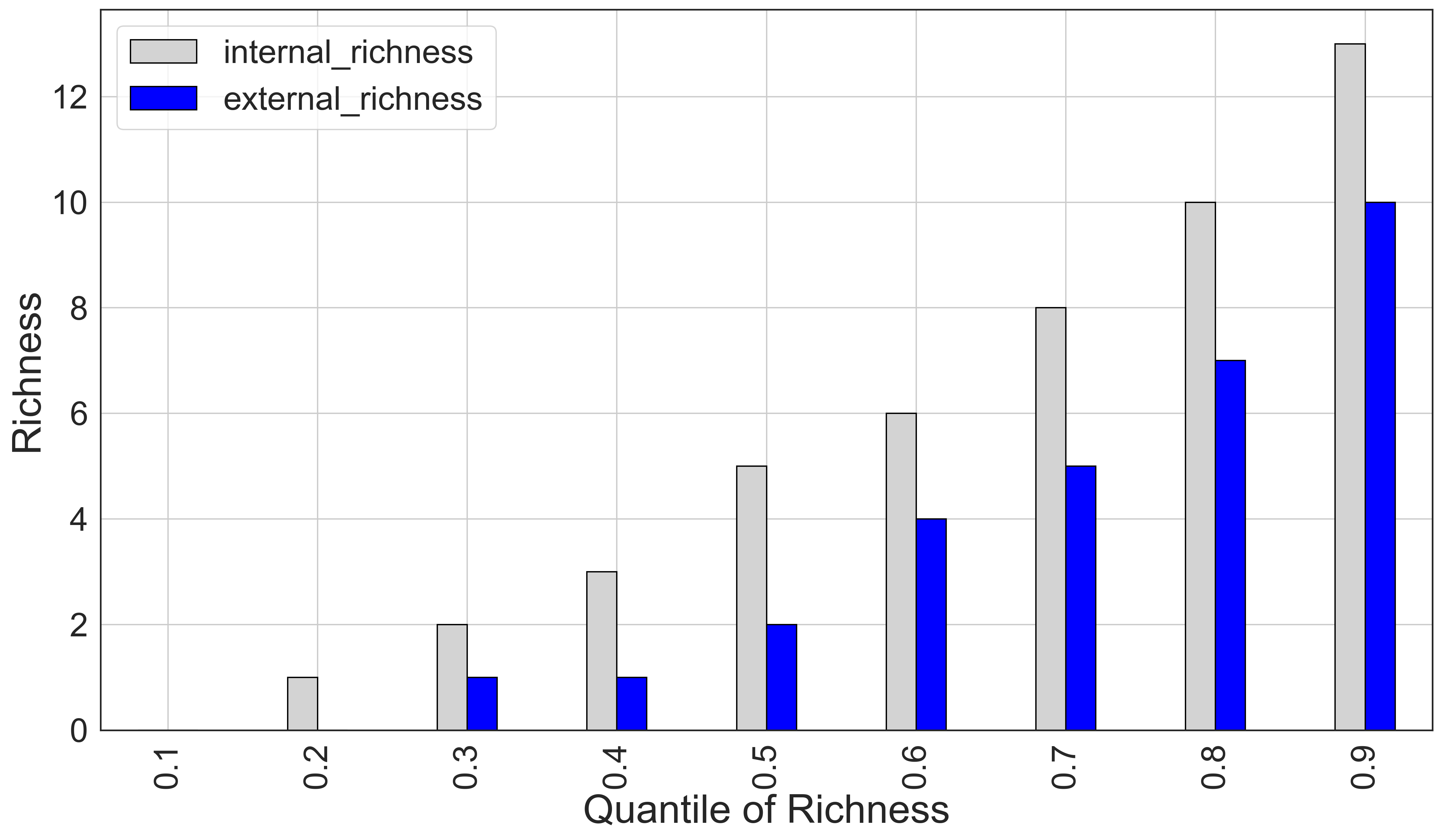}
    \caption{Comparison of Internal and External road segments in richness}
    \label{fig:galaxy}
\end{figure}

We divided the richness values of the internal and external roads from small to large into 10 equal parts and compare the split points one by one, Fig. 11 shows that the richness value of the internal roads is greater than the value of external roads at every split quantile point, indicating that for the overall road network structure, the closer to the center of the road network, the more types of functional aggregations there will be. To further explain it, as the road networks expand outward around the center gradually, the types of functions will gradually decrease.

\begin{figure}[ht]%调节图片位置，h：浮动；t：顶部；b:底部；p：当前位置
    \centering
    \includegraphics[width=0.95\textwidth]{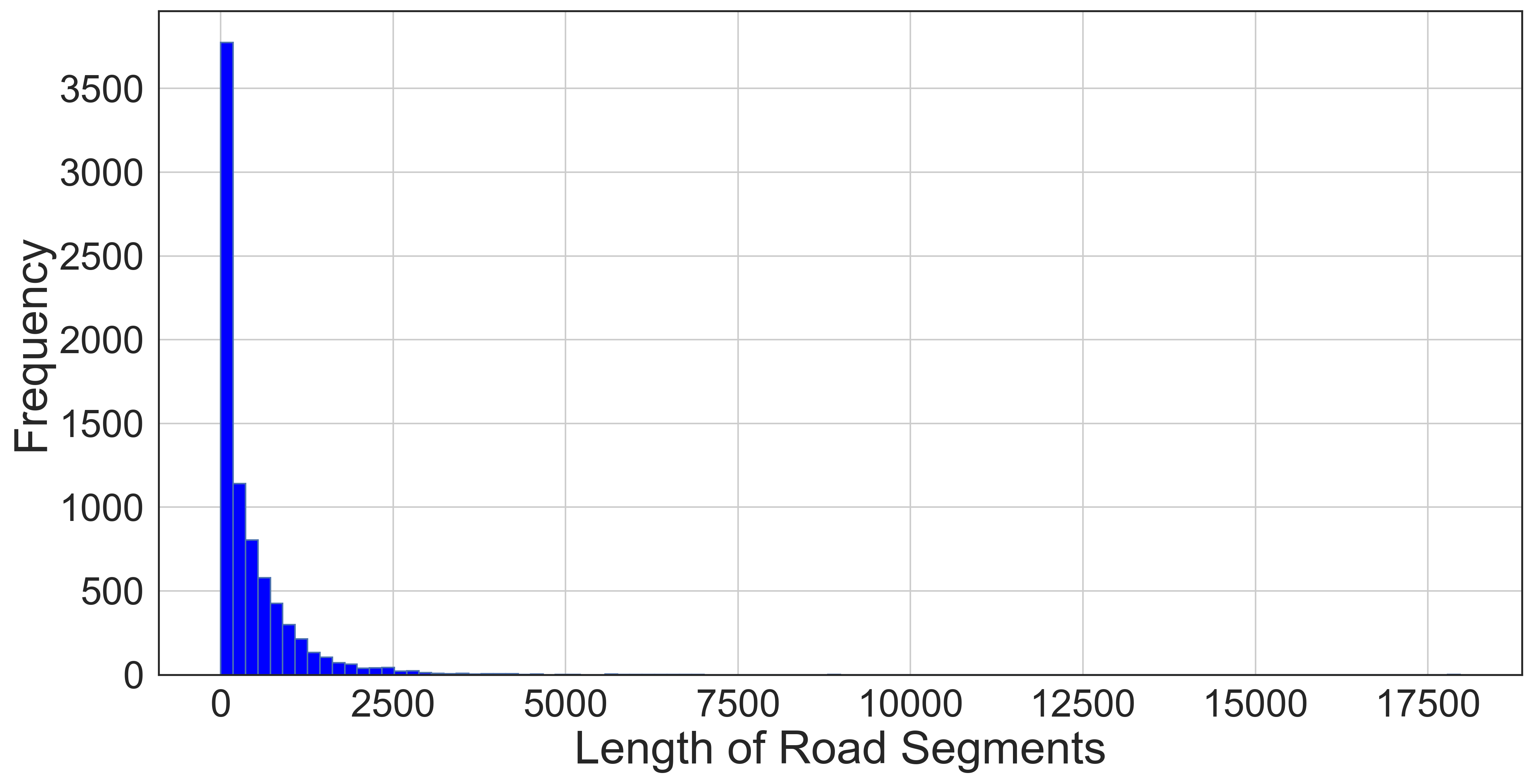}
    \caption{Length-frequency of road segments}
    \label{fig:galaxy}
\end{figure}

For length and betweenness, the Fig. 9(b) and Fig. 9(c) do not show an apparent pattern of numerical changes. What can be clearly observed is that the corresponding values of these two features both concentrate around the smaller side of the value range. Betweenness represents the 27
possible traffic flow on each road segment and Fig. 9(b) shows that it fluctuates around a certain value of richness without any certain pattern. The length directly represents the physical length of each road segment and Fig.12 shows most road segments’ length is around 1000 meters and below. Since the length of appropriate street is usually around 300-1000 meters (Yigitcanlar et al.,2007; Aklilu et al.,2018), so we further confirm that the small-scale grids are more conducive to mixed functions

\subsubsection{Interaction patterns of 4 road network substructure and mixed functions}

Because of the characteristics of unsupervised learning, the original data set has no label, so the number of clusters does not have a fixed value. Researchers usually determine the specific number of clusters based on some indicators or actual experience. In this section, we cluster our road networks to 4 categories combining the elbow method. Elbow method, also called “jump” method, determines the number of clusters by minimizing error on the basis of information theory (“Determining the number of clusters in a dataset”). Here we use the distortion of clusters as the standard for choosing the best number of clusters in the elbow method. Fig. 13(a) shows that the value of number of clusters at the “elbow” point after which begins decreasing linearly is 4. And after checking with the actual situation, we cluster the road network structure into 4 sub-structures and the richness distribution of these 4 sub-structures and the richness distribution of these sub-structures is shown in Fig.13(b). Fig.(14) display the cluster results of  
road networks and every road network substructure is annotated as follows:
\begin{figure}[ht]%调节图片位置，h：浮动；t：顶部；b:底部；p：当前位置
    \centering
    \includegraphics[width=0.95\textwidth]{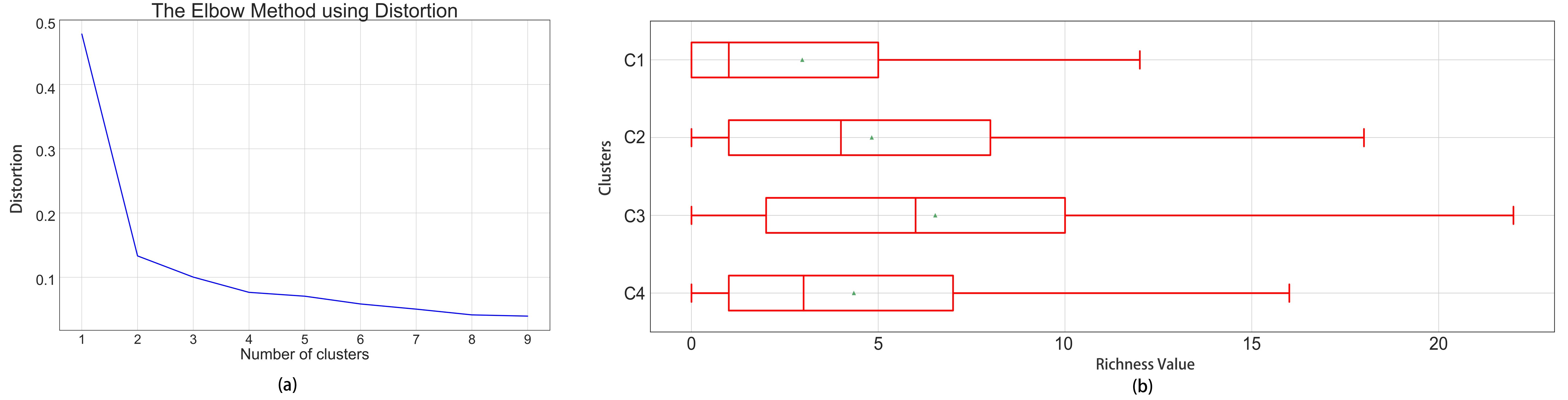}
    \caption{(a) Distortion of different cluster numbers; (b) Distribution of 4 clusters’ richness}
    \label{fig:galaxy}
\end{figure}

\begin{figure}[ht]%调节图片位置，h：浮动；t：顶部；b:底部；p：当前位置
    \centering
    \includegraphics[width=0.95\textwidth]{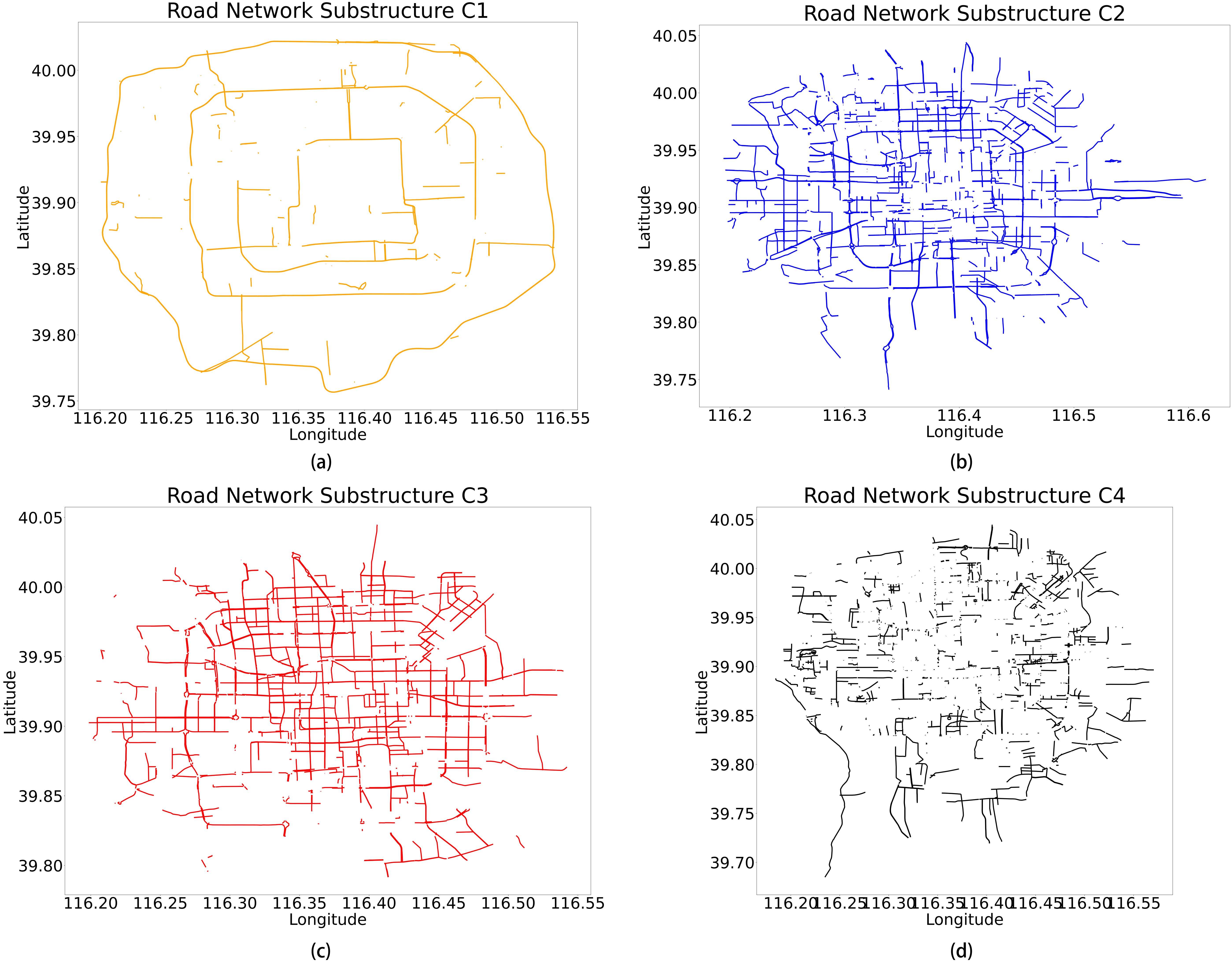}
    \caption{Richness distribution of 4 road network sub-structures}
    \label{fig:galaxy}
\end{figure}

\textbf{Main Arterial roads} C1:In this road network substructure, most road segments are urban arterial roads or urban ring roads, which carry most of the city’s traffic load and also constitute the skeleton of urban spatial form. From Fig. 13(b), we can observe that the average richness of road segments in C1 is about 10, which is way less than C2 and C3

\textbf{Mixed Secondary Arterial Roads and Branch Roads C2}: In this road network substructure, road segments are mainly composed of urban secondary roads and branch roads. The urban secondary roads usually take the part of connecting urban main roads and sharing traffic from it. Branch roads are roads connecting secondary arterial roads and streets to solve local traffic problems and focus on service functions. The distribution of richness of C2 and C3 is quite similar and obviously bigger than that of C1 and C4, indicating that mixed urban secondary arterial roads and branch roads is more to the benefit of attracting functions aggregation. After all some of the main branch roads can supplement the shortage of the main road networks and serve as public transportation lines can be set up, and can also be used as non-motor vehicle dedicated roads.

\textbf{Main Secondary Arterial Roads C3}: In this road network substructure, road segments are basically urban secondary roads, which are the same type of secondary road as in C2. However, the obvious difference between C3 road network substructure and C2 road network substructure is that C3 is basically composed of urban secondary arterial roads, while C2 is composed of both urban secondary arterial roads and branch roads, and the ratio of these 2 kinds of road segments is not particularly disparate. Fig. 13(b) shows that the mean of richness of C3 is about 6 while that of C2 is about 4, indicating that secondary arterial roads attract functions better that branch roads.

\textbf{Roads Below Branch Roads C4}: In this road network substructure, road segments are mainly roads below branch roads which are generally auxiliary roads and do not affect the urban structure or traffic function. Usually, roads like these do not attract functions aggregation a lot, but they maybe some supplementary roads for other kind of roads sometimes, so we can see from Fig. 13(b) that the mean of richness of C4 is around 3, smaller than C3, but close to C2, bigger than C1.

However, in order to see why C3 has the biggest richness than other road network structure and how features affect the aggregation of mixed functions correspondingly, we make a comparison of features of each road network structure. Because of the similarity of distribution of richness, we compare C1 and C4 together, C2 and C3 together as shown in Fig. 15 and Fig. 16.

\begin{figure}[p]%调节图片位置，h：浮动；t：顶部；b:底部；p：当前位置
    \centering
    \begin{minipage}{0.95\textwidth}
        \centering
        \includegraphics[width=0.95\textwidth]{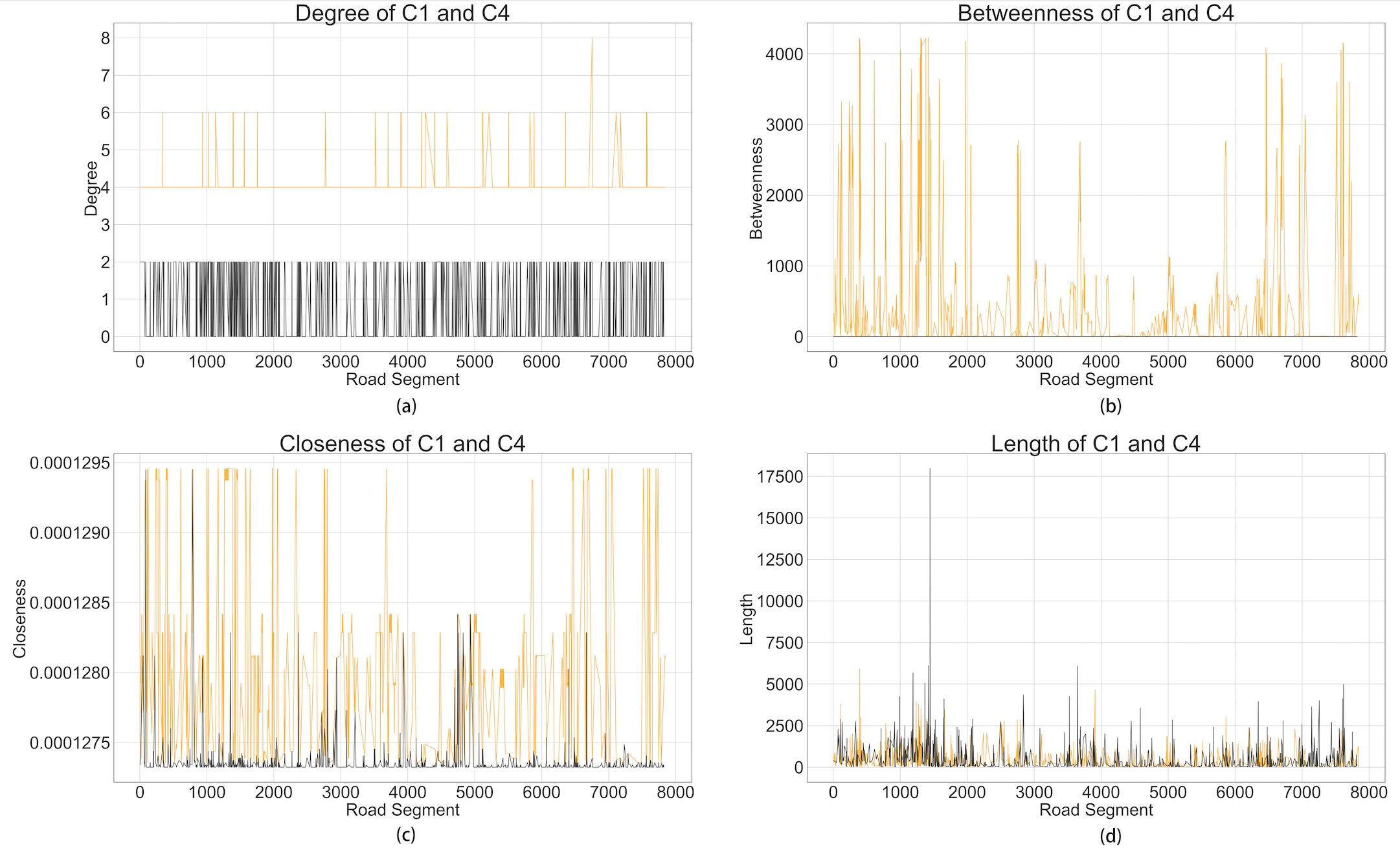}
        \caption{Distribution of 4 features of C1 and C4}
        \label{fig:galaxy}
    \end{minipage}
    
    \qquad
	%让图片换行
	\begin{minipage}{0.95\textwidth}
        \centering
        \includegraphics[width=0.95\textwidth]{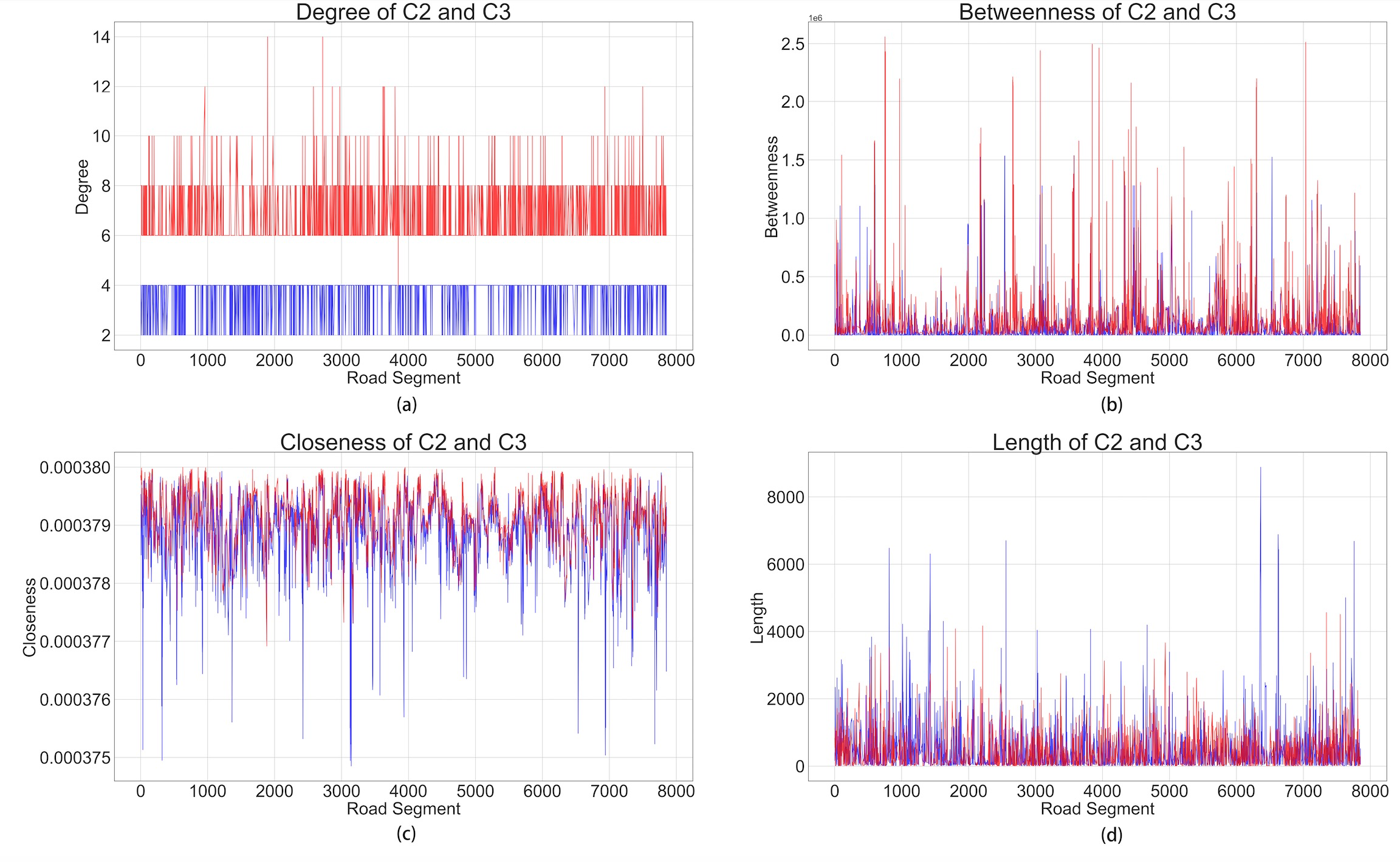}
        \caption{Distribution of 4 features of C2 and C3}
        \label{fig:galaxy}
    \end{minipage}

\end{figure}

Through Fig. 15 and Fig. 16, we can clearly find that there is a significant difference in the two features of betweenness and closeness between C1\&C4 and C2\&C3. The values of C1 and C4 on betweenness range from 0 to 4000, and the values on closeness range from 1.275 * 1e-4 to 1.295 * 1e-4. The value range of C2 and C3 in betweenness is 0 to 2.5 * 1e6, and the value range in closeness is between 3.75* 1e-4 and 3.80 * 1e-4.Betweenness represents the traffic load based on the shortest path principle, and closeness represents the distance from the center of the road network. Therefore, the difference in the value range of these features further illustrates that the common characteristics of road network substructure of C2 and C3 is closer to the center of the road network and has both transportation and life service functions, which is better for attracting functions' aggregation (Fig.13(b)), while the main function of C1 and C4 is to undertake transportation load(C1) or auxiliary functions(C4) and naturally cannot gather a large number of functions.

Although we have verified that the structure of C1 and C4 cannot attract a large number of functions' aggregation, the overall richness of C4 is still higher than that of C1. For C1 and C4, except for the feature of length, they differ a lot in the other three features. Because C1 are main arterial roads, so it will naturally have more traffic load than C4 which are mainly roads below branch roads.And Fig. 15 indeed shows that C1 has greater betweenness than C4 and also fluctuates more. For the degree, Because C4 has a lot of auxiliary roads, so it's normal that they connect few road segments and C1's roads connection with more road segments agree with its main traffic-carrier function. And this is probably the most likely explanation for why C4 has greater richness than C1. Besides, for the closeness, both C1 and C4 fluctuate a lot and don't show any exact patterns.

As for C2 and C3, Fig. 16 shows that the whole change patterns of betweenness, closeness and length of these 2 road network substructures are basically the same. This is not surprising since their structure are pretty alike as shown in Fig. 14. But C3 clearly has greater richness than C2 even though they have similar features. And the only obvious difference we can see from Fig. 16 is the degree.C2's degree range stably between 2 and 4 while C3's degree ranges from 6 to 12 and has more fluctuations. As we have found in the last section that the higher the degree is, the more mixed functions the road networks will attract in a certain range of values of degree, we now further find out that the diverse values of degree also contributes to aggregation of mixed functions. To be more specific, if2 road network structure are similar, then the one that has more ways of connection, indicating has a large value of degree, will attract more mixed functions.

% \begin{figure}[p]%调节图片位置，h：浮动；t：顶部；b:底部；p：当前位置
%     \centering
%     \includegraphics[width=0.95\textwidth]{c2c3.png}
%     \caption{Distribution of 4 features of C2 and C3}
%     \label{fig:galaxy}
% \end{figure}

\section{Conclusions}

In this paper, we propose a new framework to discover the specific interaction patterns of the road network structure and mixed functions. Through xgboost and K-means, we analyzed the overall structure and sub-structures of the road network, and found their different patterns with mixed functions respectively. From the perspective of overall structure, our analysis results show that small-scale grids and approach to the geometric center of road networks help to promote mixed functions greatly. More specific, the greater the degree and closeness of the road segments are, the more possible it will attract the aggregation of mixed functions. From the perspective of sub-structures, it is mainly the mix of urban secondary arterial roads and branch roads that contribute to mixed functions, and the main arterial road segments or roads below the ranch are not suitable for too many functions.Generally speaking, the patterns we extract from this research can not only help urban planners create a more vibrant city but also has lots of other applications like commercial site selection, social recommendations and so on.

However, there are also some limitations in this research. In the experiment, we used the number of POIs types within 100 meters around each road segment to measure the richness of mixed functions. Although 100 meters is a distance range that is closer to the road segment as well as contain most POIs demanded for calculation, the richness of different buffer distances may be different and xgboost's performance in this experiment might as well be further optimized. In future's work, in addition to solving the above problems, we will expand our framework to more cities to see if we can get a consistent conclusion at a larger spatial scale and find more interaction patterns.

\section{Acknowledgments}
This study was supported by the Key Special Project of International Science and Technology Innovation Cooperation of the National Ministry of Science and Technology (Grant No. 55 2017YFE0118600).

\bibliographystyle{alpha}
\bibliography{main}

\newcommand{\etalchar}[1]{$^{#1}$}
\begin{thebibliography}{HPH{\etalchar{+}}93b}

\bibitem[ARU19]{15}
Farhad Ahmadzai, KM~Lakshmana Rao, and Shahzada Ulfat.
\newblock Assessment and modelling of urban road networks using integrated
  graph of natural road network (a gis-based approach).
\newblock {\em Journal of Urban Management}, 8(1):109--125, 2019.

\bibitem[Bat09]{16}
Michael Batty.
\newblock Accessibility: in search of a unified theory, 2009.

\bibitem[CG16]{29}
Tianqi Chen and Carlos Guestrin.
\newblock Xgboost: A scalable tree boosting system.
\newblock In {\em Proceedings of the 22nd acm sigkdd international conference
  on knowledge discovery and data mining}, pages 785--794, 2016.

\bibitem[FP94]{10}
Lawrence~D Frank and Gary Pivo.
\newblock Impacts of mixed use and density on utilization of three modes of
  travel: single-occupant vehicle, transit, and walking.
\newblock {\em Transportation research record}, 1466:44--52, 1994.

\bibitem[Geh87]{12}
Jan Gehl.
\newblock {\em Life between buildings}, volume~23.
\newblock New York: Van Nostrand Reinhold, 1987.

\bibitem[GKR12]{20}
Karst~T Geurs, Kevin~J Krizek, and Aura Reggiani.
\newblock {\em Accessibility analysis and transport planning: challenges for
  Europe and North America}.
\newblock Edward Elgar Publishing, 2012.

\bibitem[GM80]{3}
David~M Grether and Peter Mieszkowski.
\newblock The effects of nonresidential land uses on the prices of adjacent
  housing: Some estimates of proximity effects.
\newblock {\em Journal of Urban Economics}, 8(1):1--15, 1980.

\bibitem[GWB97]{4}
Jacqueline Geoghegan, Lisa~A Wainger, and Nancy~E Bockstael.
\newblock Spatial landscape indices in a hedonic framework: an ecological
  economics analysis using gis.
\newblock {\em Ecological economics}, 23(3):251--264, 1997.

\bibitem[HNN{\etalchar{+}}19]{27}
S~Hadri, Youssef Naitmalek, Mehdi Najib, Mohamed Bakhouya, Youssef Fakhri, and
  M~Elaroussi.
\newblock A comparative study of predictive approaches for load forecasting in
  smart buildings.
\newblock {\em Procedia Computer Science}, 160:173--180, 2019.

\bibitem[HPH{\etalchar{+}}93a]{13}
Bill Hillier, Alan Penn, Julienne Hanson, Tadeusz Grajewski, and Jianming Xu.
\newblock Natural movement: or, configuration and attraction in urban
  pedestrian movement.
\newblock {\em Environment and Planning B: planning and design}, 20(1):29--66,
  1993.

\bibitem[HPH{\etalchar{+}}93b]{26}
Bill Hillier, Alan Penn, Julienne Hanson, Tadeusz Grajewski, and Jianming Xu.
\newblock Natural movement: or, configuration and attraction in urban
  pedestrian movement.
\newblock {\em Environment and Planning B: planning and design}, 20(1):29--66,
  1993.

\bibitem[HZGH08]{31}
Guobiao Hu, Shuigeng Zhou, Jihong Guan, and Xiaohua Hu.
\newblock Towards effective document clustering: A constrained k-means based
  approach.
\newblock {\em Information Processing \& Management}, 44(4):1397--1409, 2008.

\bibitem[Jac61]{2}
Jane Jacobs.
\newblock Jane jacobs.
\newblock {\em The Death and Life of Great American Cities}, 21(1):13--25,
  1961.

\bibitem[Jos06]{11}
Lou Jost.
\newblock Entropy and diversity.
\newblock {\em Oikos}, 113(2):363--375, 2006.

\bibitem[Kar12]{25}
Kayvan Karimi.
\newblock A configurational approach to analytical urban design:‘space
  syntax’methodology.
\newblock {\em Urban Design International}, 17(4):297--318, 2012.

\bibitem[KR12]{6}
Hans~RA Koster and Jan Rouwendal.
\newblock The impact of mixed land use on residential property values.
\newblock {\em Journal of regional science}, 52(5):733--761, 2012.

\bibitem[LGL19]{22}
Kang Liu, Song Gao, and Feng Lu.
\newblock Identifying spatial interaction patterns of vehicle movements on
  urban road networks by topic modelling.
\newblock {\em Computers, Environment and Urban Systems}, 74:50--61, 2019.

\bibitem[LLD{\etalchar{+}}18]{14}
Feng Lu, Kang Liu, Yingying Duan, Shifen Cheng, and Fei Du.
\newblock Modeling the heterogeneous traffic correlations in urban road systems
  using traffic-enhanced community detection approach.
\newblock {\em Physica A: Statistical Mechanics and its Applications},
  501:227--237, 2018.

\bibitem[LWH11]{30}
Hui-Lan Luo, Hui Wei, and Fan-Xing Hu.
\newblock Improvements in image categorization using codebook ensembles.
\newblock {\em Image and Vision Computing}, 29(11):759--773, 2011.

\bibitem[PT04]{17}
Alan Penn and Alasdair Turner.
\newblock Movement-generated land-use agglomeration: simulation experiments on
  the drivers of fine-scale land-use patterning.
\newblock {\em Urban Design International}, 9(2):81--96, 2004.

\bibitem[Rat04]{19}
Carlo Ratti.
\newblock Space syntax: some inconsistencies.
\newblock {\em Environment and Planning B: Planning and Design},
  31(4):487--499, 2004.

\bibitem[RCS{\etalchar{+}}19]{28}
N~Reynaldo, William Chanrico, Derwin Suhartono, Fredy Purnomo, et~al.
\newblock Gender demography classification on instagram based on user’s
  comments section.
\newblock {\em Procedia Computer Science}, 157:64--71, 2019.

\bibitem[SK04]{5}
Yan Song and Gerrit-Jan Knaap.
\newblock Measuring the effects of mixed land uses on housing values.
\newblock {\em Regional Science and Urban Economics}, 34(6):663--680, 2004.

\bibitem[SK16]{21}
Yao Shen and Kayvan Karimi.
\newblock Urban function connectivity: Characterisation of functional urban
  streets with social media check-in data.
\newblock {\em Cities}, 55:9--21, 2016.

\bibitem[SKX13]{18}
Y~Shen, K~Karimi, and Q~Xia.
\newblock Morphological transformation of historical centres in tianjin.
\newblock In {\em 2013 International Space Syntax Symposium}, 2013.

\bibitem[SP03]{1}
Dean Schwanke and Patrick~L Phillips.
\newblock {\em Mixed-use development handbook}.
\newblock Urban Land Inst, 2003.

\bibitem[WCMZ20]{23}
Mingshu Wang, Zheyan Chen, Lan Mu, and Xuan Zhang.
\newblock Road network structure and ride-sharing accessibility: A network
  science perspective.
\newblock {\em Computers, environment and urban systems}, 80:101430, 2020.

\bibitem[YHTC19]{24}
Yuanxuan Yang, Alison Heppenstall, Andy Turner, and Alexis Comber.
\newblock A spatiotemporal and graph-based analysis of dockless bike sharing
  patterns to understand urban flows over the last mile.
\newblock {\em Computers, Environment and Urban Systems}, 77:101361, 2019.

\bibitem[YWW20]{9}
Feng Yuan, Yehua~Dennis Wei, and Jiawei Wu.
\newblock Amenity effects of urban facilities on housing prices in china:
  Accessibility, scarcity, and urban spaces.
\newblock {\em Cities}, 96:102433, 2020.

\bibitem[YZY{\etalchar{+}}17]{7}
Yang Yue, Yan Zhuang, Anthony~GO Yeh, Jin-Yun Xie, Cheng-Lin Ma, and Qing-Quan
  Li.
\newblock Measurements of poi-based mixed use and their relationships with
  neighbourhood vibrancy.
\newblock {\em International Journal of Geographical Information Science},
  31(4):658--675, 2017.

\bibitem[ZLT16]{8}
Yinan Zhou, Ziyue Li, and Xinyu Tao.
\newblock Urban mixed use and its impact on energy performance of micro gird
  system.
\newblock {\em Energy Procedia}, 103:339--344, 2016.

\end{thebibliography}

\end{document}